\documentclass[12pt]{article}
\usepackage{graphicx}
\usepackage{amsmath,amssymb}
\usepackage{hyperref}
\begin{document}
\def\be{\begin{equation}}
\def\ee{\end{equation}}
\def\ba{\begin{eqnarray}}
\def\ea{\end{eqnarray}}	
\def\l{\left}
\def\r{\right}
\def\fr{\frac}
\def\la{\label}
\def\d{\partial}
\def\vphi{\varphi}
\def\pr{\prime}
\def\mpl{M_{\rm p}}
\def\hub{\mathcal{H}}
\def\mbf{\mathbf}
\def\baselinestretch{1.2}


\author{Bart Horn\thanks{bhorn01@manhattan.edu}}
\title{\bf\Large The Higgs field and early universe cosmology: a (brief) review}
\date{\normalsize\textit{Manhattan College, Department of Physics,\\ New York, NY 10471, USA}}

{\let\newpage\relax\maketitle}
\thispagestyle{empty}
\vskip 0.5cm
\begin{abstract}
\noindent We review and discuss recent work exploring the implications of the Higgs field for early universe cosmology, and vice versa.  Depending on the model under consideration the Higgs may be one of a few scalar fields determining the evolution and fate of the Universe, or the Higgs field may be connected to a rich sector of scalar moduli with complicated dynamics.  In particular we look at the potential consequences of the Higgs field for inflation and its predictions, for the (meta)stability of the Standard Model vacuum, and for the existence of dynamical selection mechanisms in the landscape.
\end{abstract}




\section{Introduction: implications of $m_h$ = 125 GeV}

The observation of the Higgs boson in 2012 \cite{Aad:2012tfa, Chatrchyan:2012xdj}, with a mass of 125 GeV, marked the last piece of the Standard Model (SM) of particle physics to be discovered and inaugurated a new era of precision Higgs boson physics.  Further experiments have measured the properties and couplings of the Higgs boson to increasing levels of precision and confirmed the consistency of the discovered particle with the predicted properties of the SM Higgs boson \cite{Aaboud:2017vzb, Sirunyan:2018koj, Aad:2019mbh}.  Ongoing and planned experiments at the energy and intensity frontier \cite{Arkani-Hamed:2015vfh, ApollinariG.:2017ojx, Charles:2018vfv, Abada:2019lih} will measure the Higgs boson couplings to greater levels of precision and test the predictions of the SM at higher energies.

Despite the many recent successes, however, many deep questions remain.  So far the Large Hadron Collider has not observed any evidence for weak-scale supersymmetry or other particles beyond the Standard Model, leaving the physics that sets the scale of the Higgs boson mass  unresolved.  The ``hierarchy problem'' consists of the observation that the Higgs boson mass is potentially sensitive to quantum corrections, and the bare (unrenormalized) Higgs mass may therefore need to be fine-tuned in order for the physical Higgs boson mass to end up many orders of magnitude smaller than the Planck scale.  While many different mechanisms (such as weak-scale supersymmetry) have been proposed to resolve the hierarchy problem, even in the context of these proposed solutions, the observed value of $m_h = 125$ GeV does not clearly favor either a dynamical or an anthropic selection mechanism for the exact value of the Higgs boson mass, nor whether this value is natural or fine-tuned (see e.g. \cite{Dine:2015xga}).  Continuing experiments at higher energies and intensities may help to resolve these questions; however, the absence of data indicating the existence of new physics at higher scales poses challenges for the design and implementation of the next generation of collider searches.
In this situation a convergence between particle physics and cosmology may be best opportunity to probe or constrain model building, given that extremely high energies have already been achieved by the conditions in the early universe.  In this brief review, we will discuss consequences of early universe cosmology for the physics of the Higgs field, and vice versa.

While an energy desert between the electroweak and Planck scales, if confirmed, may be disappointing from the point of view of particle physics and model building, it also raises the possibility that the low-energy Standard Model continues to apply at very high energy scales, providing direct connections between low and high energy physics.  In particular the Higgs field may be connected to the physics of inflation and reheating, with only minimal modifications to the Standard Model required \cite{Bezrukov:2007ep}.  Other studies indicate that the measured values of the Higgs and top mass, if continued via renormalization group (RG) running to high energies, place the Standard Model on the edge of metastability \cite{Degrassi:2012ry, Buttazzo:2013uya}.  Measurements at low energies (i.e., accessible at LHC scales and slightly above) energies may therefore be of great significance for understanding the past and future evolution of our Universe. 

So far the Higgs field is the only known candidate fundamental scalar field observed in nature.  Scalar fields are ubiquitous in models of primordial cosmology, however, especially in models of inflation and (p)reheating (see e.g. \cite{Guth:1980zm, Linde:1981mu, Albrecht:1982wi, Linde:1983gd, Kofman:1997yn}), and theories with many scalar moduli fields can arise (and may even be required) as a generic prediction of string compactification and supersymmetric field theory models.  These additional scalar moduli and their associated field condensates may have important consequences for models of cosmology during and after the era of inflation, and may even affect the standard cosmic history \cite{Coughlan:1983ci, deCarlos:1993wie, Banks:1993en}.  If the Higgs potential is only a small piece of a larger scalar sector, we might hope that an understanding of the Higgs field and its dynamics at large field ranges can help us to probe these additional fields and learn about the broader landscape.  

The literature on the Higgs field and cosmology is quite extensive, and this brief review is intended to serve as an introduction to a large and rapidly developing field.  (For prior reviews at the intersection of Higgs physics and cosmology, see e.g. \cite{Bezrukov:2015ytd, Shaposhnikov:2015mja, Moss:2015fma, Rajantie:2018tqm, Rubio:2018ogq, Steinwachs:2019hdr}.)  The references provided are intended to be illustrative, but it should be emphasized that they are by no means exhaustive.  This review is organized as follows: in \S 2 we discuss the Higgs field and inflation, starting with models where the Higgs field itself serves as the inflaton, and continuing to discuss models where the Higgs field affects the generation of primordial perturbations as a spectator field.  In \S 3 we discuss the consequences of the Higgs boson mass and couplings for the stability of the Standard Model vacuum. In \S 4 we discuss the Higgs field and selection mechanisms in the landscape, paying particular attention to dynamical selection mechanisms inspired by studies of nonperturbative moduli dynamics.  In \S 5 we conclude and indicate further directions.

\section{The Higgs field and inflation}

Inflation sourced by the energy of a slowly rolling scalar field can solve the flatness and horizon puzzles in cosmology \cite{Guth:1980zm, Linde:1981mu, Albrecht:1982wi, Linde:1983gd}, as well as provide a mechanism to generate the primordial perturbations which form the seeds of structure \cite{Mukhanov:1981xt} (see e.g. \cite{Baumann:2014nda} for a review).  Given that the Higgs is the only known candidate fundamental scalar field in nature, it is of interest to explore whether it is possible for it to serve as the inflaton.  In a framework with many scalar fields (such as may arise in string compactifications and in supersymmetric field theories) there is no particular necessity for the Higgs and the inflaton to be the same field; however, given that the Higgs is well understood at low energies it also serves as a useful model for studying the behavior of scalar fields during inflation and shortly after the end of inflation.  In the subsection below we review the status of models where the Higgs itself is the inflaton, and in the following subsection we discuss models where the Higgs affects the growth of curvature perturbations and structure as a spectator field.


\subsection{The Higgs field as the inflaton}

The Higgs sector action in the Standard Model is given by
\begin{equation}\label{HiggsPotentialSM}
S = \int d^{4}x \sqrt{-g} \left(D_{\mu}\mathcal{H}^{\dagger}D^{\mu}\mathcal{H} + \mu^2 \mathcal{H}^{\dagger}\mathcal{H} - \lambda (\mathcal{H}^{\dagger}\mathcal{H})^2\right)
\end{equation}
where $\mathcal{H}$ is a doublet under the $SU(2)$ gauge group of the weak interactions\footnote{Throughout this paper we will use $\mathcal{H}$ to denote the Higgs doublet field, $h$ for the field corresponding to the physical Higgs boson, $h_{\mu \nu}$ for the canonically normalized graviton field in flat space, and $H$ to denote the Hubble parameter.}.  The term $\mu^2 > 0$ causes the symmetry to spontaneously break at low energies, and fluctuations around the vacuum expectation value (vev) are parameterized by
\begin{equation}
\mathcal{H} = \frac{1}{\sqrt{2}}\left({\begin{array}{c}
   0 \\
   v + h  \\
  \end{array} }\right)\,,
\end{equation}
where $v = 246$ GeV is the scalar field vev, $h$ is the scalar Higgs boson with mass $m_{h} = \sqrt{2}\mu = \sqrt{2\lambda}v$, and we have chosen to work in unitary gauge, in which case the three additional Goldstone boson degrees of freedom are eaten by the massive vector bosons.  At large field values $h \gg v$ the potential energy is dominated by the quartic term,
\begin{equation}
S \approx \int d^{4}x \sqrt{-g}\left(\frac{1}{2}\partial_{\mu}h \partial^{\mu}h - \frac{\lambda}{4}h^4\right)\,.
\end{equation}
It is straightforward to show (see e.g.\ \cite{Linde:1983gd}) that this potential cannot source slow-roll inflation and produce fluctuations consistent with the observed inflationary power spectrum for $\lambda$ near its low-energy value $\lambda \approx 0.13$; a value closer to $\lambda \sim 10^{-13}$ would be required.  It was shown in \cite{Bezrukov:2007ep} (see also e.g. \cite{Salopek:1988qh, Fakir:1990eg} for previous models of chaotic inflation featuring this term) that the Standard Model Higgs can nevertheless serve as the inflaton with the inclusion of the term
\begin{equation}
\mathcal{L} \supset \xi \mathcal{H}^{\dagger}\mathcal{H} R
\end{equation}
This is a renormalizable term which can arise naturally in the Standard Model coupled to gravity, and may be measurable (or at least constrainable) in environments with large spacetime curvature.  

The Higgs field action with gravity at large field values $h \gg v$ then becomes
\begin{equation}
S \approx \int d^{4}x \sqrt{-g}\left(\frac{M_P^2}{2} R + \frac{\xi}{2}h^2 R + \frac{1}{2}\partial_{\mu}h \partial^{\mu}h - \frac{\lambda}{4}h^4\right)
\end{equation}
which renormalizes the Planck mass for large field values of $h$.  Switching to the Einstein frame
\begin{equation}
g_{\mu \nu} \to g_{E\mu \nu}\left(1 + \frac{\xi h^2}{M_P^2}\right)^{-1}\,,
\end{equation}
to fix the coefficient of the gravitational kinetic term, and writing the Higgs kinetic term in Einstein frame in terms of the canonically normalized field $\phi$,
the potential at large field ranges $h \gg M_P/\sqrt{\xi}$ becomes
\begin{equation}\label{HiggsInflation}
V(\phi) \approx \frac{\lambda M_P^4}{4 \xi^2}\left(1 + e^{-\sqrt{\frac{2}{3}}\frac{\phi}{M_P}}\right)^{-2}\,.
\end{equation}
The final potential in terms of the canonically normalized field is an exponential, similar to the predictions of $R + R^2$ type models \cite{Starobinsky:1980te}.  While it is easy to find a parametric window where this potential satisfies the conditions for slow-roll inflation, a value of $\xi \approx 49000\sqrt{\lambda}$ is necessary to keep the inflationary perturbations consistent with the observed power spectrum.  The net effect of the $(\xi/2) h^2 R$ coupling is therefore to flatten the Higgs potential at large field values; here the flattening is accomplished by coupling to the gravitational field, but this type of flattening effect can arise more generally when the inflaton backreacts on heavy fields during inflation \cite{Dong:2010in}.\footnote{A related class of models is the Higgs-dilaton inflation model, which exchanges the dimensionful Planck scale $M_P$ for the vev of an extra scalar degree of freedom.  This has been used in \cite{GarciaBellido:2011de, Bezrukov:2012hx} to realize inflation and dark energy within a single model.}

It is straightforward to calculate the predicted values of inflationary precision observables, such as the scalar tilt $n_s$, the tensor-to-scalar ratio $r$ and the non-Gaussianities (see e.g. \cite{Baumann:2014nda} for precise definitions of these quantities), given the form of the inflationary potential in Equation \eqref{HiggsInflation}: in \cite{Bezrukov:2007ep} a few of these values are given as 
\begin{equation}\label{HiggsInflationPrecisionObservables}
n_s \approx 0.97\,, \qquad \qquad r = 0.0033\,.  
\end{equation}
These values are consistent with current observations \cite{Aghanim:2018eyx}; however, even if these precise values are confirmed by experiment some extra information may be required in order to provide evidence that such a potential corresponds to the Standard Model Higgs boson, and an understanding of the reheating sector and corresponding cosmological history of this epoch may be relevant as well. 
Precision observables in this class of models can also be affected by non-renormalizable corrections between the electroweak scale and inflationary energies \cite{Enckell:2016xse}, and the allowed parameter window can be increased further by including corrections such as $R^2$ terms that alter the form of the inflationary potential from that given in Equation \eqref{HiggsInflation}\cite{Enckell:2018uic}.  

A more fundamental problem for this class of models concerns the consistency of the model in an effective field theory (EFT) approximation: in \cite{Burgess:2009ea, Barbon:2009ya, Lerner:2009na, Atkins:2010re, Bezrukov:2010jz} it was noted that the Hubble scale during inflation $H_{inf} \sim \frac{\sqrt{\lambda}M_P}{\xi}$ is close to the EFT cutoff for the low-energy theory for this particular class of models.  Expanding the metric $g_{\mu \nu} = \eta_{\mu \nu} + h_{\mu \nu}/M_P$ around flat space, the term $\xi h^2 R$ in the Lagrangian gives rise to graviton-Higgs couplings that can cause unitarity of the EFT to break down at a cutoff scale $M_P/\xi$.  Similar issues arise in different models of Higgs inflation, such as when a coupling of the Higgs field kinetic term to the Einstein tensor $G^{\mu \nu}D_{\mu}\mathcal{H}^{\dagger}D_{\nu}\mathcal{H}$ is used instead to source inflation \cite{Germani:2010gm, Atkins:2010yg}, and nonminimal couplings in the Higgs sector or couplings of the Higgs field to additional sectors may also give rise to unitarity problems \cite{Burgess:2010zq, Hertzberg:2010dc}.  
In was noted in \cite{Bezrukov:2010jz} that the renormalization of the Planck constant at large $h$ will make the cutoff scale field-dependent, which may help with consistency during the inflationary era.  However, connecting the effective theory that is valid during inflation to the low-energy EFT may depend the details of the UV completion and the accompanying nonrenormalizable corrections, (see also e.g. \cite{Giudice:2010ka, Barbon:2015fla}), and both the form of the potential at high energies and the corresponding predictions for precision observables in this class of models may depend sensitively on these corrections as well (see e.g. \cite{Enckell:2018kkc}).  As in the discussion in the paragraph above, these corrections may help relax the predictions for precision observables from the values in Equation \eqref{HiggsInflationPrecisionObservables}.

An understanding of the reheating sector and corresponding cosmological history of this epoch may be relevant as well in order to investigate the role of the Higgs field in inflation.  Since the couplings of the Higgs field to the rest of the Standard Model are known, an advantage of using the SM Higgs field as the inflaton is that the dynamics of reheating to the Standard Model can in principle be calculated in detail.  In \cite{GarciaBellido:2008ab, Figueroa:2009jw} it was shown that the dynamics of SM gauge bosons have important effects on the Higgs decay, with perturbative decays of the gauge bosons into fermions initially delaying or blocking the formation of a parametric resonance, and backreaction of the gauge bosons on the Higgs condensate eventually becoming important as well.
The full dynamics of reheating through the eventual formation of the resonance and eventual thermalization of the Standard Model degrees of freedom need to be solved nonperturbatively, but can in principle lead to very precise predictions.  

\subsection{The Higgs as a spectator field}

Even when the Higgs field is subdominant to the potential energy of a different inflaton field, it can undergo complex dynamics during and just after inflation, and these dynamics may affect the generation and evolution of the inflationary perturbations.  Several different mechanisms have been explored, depending on whether the Higgs field plays a leading or subleading role in the generation of perturbations.

In the curvaton scenario \cite{Linde:1996gt, Enqvist:2001zp, Lyth:2001nq, Moroi:2001ct} a spectator field (or fields) during inflation persists afterwards as an oscillating condensate that comes to dominate the energy density of the Universe, with gradient and potential energy that scales like $\rho(t) \propto a(t)^{-3}$, and its primordial fluctuations become converted into adiabatic curvature perturbations before the curvaton field decays.  The Higgs was investigated as a candidate curvaton in \cite{DeSimone:2012qr, DeSimone:2012gq, Choi:2012cp}, and its perturbations can be converted to curvature perturbations through a variety of mechanisms.  Such models generally predict constraints on the inflationary Hubble scale $H_{inf}$ where the specific constraints depend on the exact conversion mechanism used, and may predict testable relations between the scale of inflation and the beta function of the Higgs (which in turn is determined by the Higgs and top quark masses) \cite{DeSimone:2012qr}.  Precision observables such as the tensor-to-scalar ratio and primordial non-Gaussianities can also help to confirm or falsify this scenario \cite{DeSimone:2012gq, Choi:2012cp, Kunimitsu:2012xx}.  Since the Standard Model Higgs potential as given in Equation \eqref{HiggsPotentialSM} is expected to be subdominant to the inflationary potential, the curvaton decay must be suppressed for a considerable time after the end of inflation in order for it to eventually dominate the curvature perturbation. 
In \cite{Enqvist:2012tc, Enqvist:2013gwf} it was shown that curvaton decay through couplings to the Higgs field can be suppressed for awhile by the coupling of the Higgs to a thermal bath of SM particle, which generates a (temporary) temperature-dependent mass term.  


Even when the energy of the Higgs field is itself too small to directly contribute to the primordial curvature perturbations, it is still possible for it to play a role in generating the inflationary perturbations.  In the modulated reheating scenario of \cite{Dvali:2003em} the inflationary perturbations are generated when the inflaton decays at the end of inflation, which the decay rate $\Gamma(h)$ being dependent on the value of the Higgs field.  The fluctuations of the Higgs field therefore determine the relative length of time inflation lasts in nearby observer patches and become imprinted in the inflationary perturbations, without the Higgs field itself needing to satisfy the slow-roll conditions.  A similar scenario involves using Higgs portal couplings to imprint inflationary perturbations from a dark sector (see for instance \cite{Kainulainen:2016vzv}). 

Regardless of whether it plays a significant role during inflation or not, the dynamics of the Higgs field itself during this era can be very complex and may lead to testable predictions.  Like all fields present during inflation, the Higgs field will undergo fluctuations with $\delta h \sim H_{inf}$, forming a condensate which will begin to oscillate when $H \approx m_{h}$ after the end of inflation.  The fact the the couplings of the Higgs field to the rest of the Standard Model are known makes it possible to investigate the subsquent behavior and decay of the condensate in some detail.  The Higgs will decay perturbatively into SM particles, and can also decay through nonperturbative particle production when the oscillations of the Higgs condensate causes the gauge boson masses to become non-adiabatic; however, the rapid decay of SM gauge bosons into fermions may delay or prevent the formation of a parametric resonance for this channel \cite{GarciaBellido:2008ab, Figueroa:2009jw, Enqvist:2013kaa}.  Backreaction of produced gauge bosons on the Higgs field mass may affect the decay process as well, and the decay of the Higgs may need to be solved numerically \cite{Enqvist:2014tta, Enqvist:2015sua}.  The final abundances of the various SM decay products will depend in detail on the couplings to the inflationary sector and the expansion of the postinflationary universe \cite{Figueroa:2015rqa, Figueroa:2016dsc}, and may be accompanied a stochastic background of high-frequency gravitational waves \cite{Figueroa:2014aya, Figueroa:2016ojl} as well.

If the Higgs is coupled to gravity (as in the Higgs inflation scenario described in the subsection above) the sudden changes in the kinetic term at the end of inflation may allow the Higgs field to decay through very rapid preheating \cite{Ema:2016dny, Sfakianakis:2018lzf}.  Whether or not this leads to a breakdown of the EFT for the Higgs field depends on what higher-dimensional operators are present and may indirectly probe the dynamics of the UV completion \cite{Hamada:2020kuy}.

Couplings between the Higgs and other scalar moduli fields during inflation may lead to complicated dynamics as well, which can become imprinted on the inflationary power spectrum if the Higgs couples to the inflaton.  In the ``Higgscitement'' scenario of \cite{Amin:2019qrx, Fan:2019udt} a model was explored where the potential in Equation \eqref{HiggsPotentialSM} becomes moduli dependent, and in particular the Higgs mass arises from a trilinear coupling
\begin{equation}
\mathcal{L} \supset \frac{\mu^2}{2}\frac{(\phi - \phi_0)}{f}h^2
\end{equation}
where $f$ is the scale of an axion-like modulus $\phi$ which oscillates rapidly in the early universe.  In this case the dynamics of the modulus $\phi$ may give rise to complicated dynamics where the Higgs field rapidly transitions between symmetry breaking and symmetry preserving phases, which may lead to production of stochastic gravitational waves, or affect the duration of inflation and reheating \cite{Amin:2019qrx}, or may produce oscillatory features which are imprinted on the primordial power spectrum \cite{Fan:2019udt} via irrelevant operators that couple the Higgs field to the inflaton.  These oscillatory imprints on the power spectrum may be quite distinctive if observed, and even if they are generated by an unknown field with similar dynamics to the Higgs field, would indicate the presence of spontaneous symmetry breaking dynamics.

\section{The Higgs field and vacuum metastability}

A general feature of models of scalar field landscapes is the possible existence of vacuum transitions between different local minima, and it is of interest to explore the consequences for our Universe if the Standard Model vacuum is metastable.  Given that the Universe is currently entering an approximate De Sitter phase it is perhaps to be expected that the present vacuum may be metastable (see e.g.\ \cite{Silverstein:2004id, Douglas:2006es}); remarkably, the measured values of the Higgs boson and top quark masses and other parameters indicate that metastability may already be visible in the Standard Model.  Cosmological phase transitions arising from vacuum instabilities in the past may also have interesting phenomenological consequences; however, there is no evidence so far that this has occurred in the cosmological history of the Standard Model.  In the subsection below we discuss the consequences of the possible metastability of the current SM vacuum, and in the following subsection we discuss the possibility of cosmological signatures from the beavhior of the Higgs field during the electroweak phase transition.

\subsection{Stability of the SM vacuum}

Studying the loop-corrected renormalization group evolution of the parameters of the SM it has been noted that the current measured values for the Higgs boson and top quark masses predict that the Higgs coupling $\lambda$ turns negative on scales above $10^{11}$ GeV or so, which may make the SM metastable (see for instance \cite{Degrassi:2012ry, Buttazzo:2013uya} and references therein).  The beta function at one loop is dominated by the quartic coupling $\lambda$ and the top Yukawa coupling $y_t$,
\begin{equation}
\beta_{\lambda} = \frac{d\lambda}{d \log \mu} = \frac{1}{(4\pi)^2}(24 \lambda^2 + 12 \lambda y_t^2 - 6 y_t^2) + \cdots
\end{equation}
where $\mu$ is the renormalization scale.  Here we have neglected gauge couplings and other Yukawa couplings, which are necessary to calculate the full beta function precisely.  In \cite{Degrassi:2012ry, Buttazzo:2013uya} the RG running of the SM parameters is calculated to two loops, and the scale $\mu_0$ where $\lambda(\mu_0)$ turns negative is given in \cite{Buttazzo:2013uya} as
\begin{equation}
\log \frac{\mu_0}{\textrm{GeV}} = 11.3 + 1.0\left(\frac{m_h}{\textrm{GeV}} - 125.66\right) - 1.2\left(\frac{m_t}{\textrm{GeV}} - 173.10\right) + 0.4\left(\frac{\alpha_{3}(m_Z) - 0.1184}{0.0007}\right)
\end{equation}
The SM appears in fact to be close to the edge of metastability, with the fate of the vacuum depending sensitively on the exact masses and couplings (see also \cite{ArkaniHamed:2008ym}).  Varying the Higgs and top quark masses by only a few GeV from their measured values would restore stability (which is ruled out at $2-3\sigma$ in \cite{Degrassi:2012ry, Buttazzo:2013uya}), motivating further precision studies of both these parameters, e.g.\ at future linear colliders.  The observation that the SM lies on the edge of criticality between stable and unstable phases may have either a statistical/anthropic explanation or a dynamical one (see also \cite{Cline:2018ebc, Khoury:2019ajl} for landscape based proposals), and is a pattern repeated elsewhere in the SM, since the smallness of the Higgs mass squared puts the SM close to the boundary between phases with broken and unbroken electroweak symmetry.  It is also noted in \cite{Degrassi:2012ry, Buttazzo:2013uya} that given the measured values of the Higgs and top quark masses, both the Higgs quartic coupling $\lambda(\mu)$ and its beta function are nearly vanishing at the Planck scale $\mu \sim M_P$; however, it is not clear whether this is set by dynamics of quantum gravity at or near the Planck scale (see e.g. \cite{Shaposhnikov:2009pv, Hebecker:2012qp, Hebecker:2013lha}) or the near-vanishing is merely a coincidence\footnote{Nevertheless, it is interesting to note that the initial conditions $\lambda(M_P) = 0$ were used to predict $m_h \approx 126$ GeV in \cite{Shaposhnikov:2009pv}, several years prior to the discovery of the Higgs boson.}.  

While the lifetime of the metastable electroweak vacuum is expected to be many times longer than the present age of the Universe, the existence of the lower-energy (``true'') vacuum provides a strong constraint on cosmological models of inflation and reheating \cite{Espinosa:2007qp, EliasMiro:2011aa, Kobakhidze:2013tn, Fairbairn:2014zia, Enqvist:2014bua, Kobakhidze:2014xda, Hook:2014uia, Espinosa:2015qea, Kohri:2016wof, Enqvist:2016mqj, Grobov:2015ooa, Jain:2019wxo} (see also \cite{Markkanen:2018pdo} for a recent review).  It is necessary to make sure that the SM vacuum is not destabilized by too high a reheating tempature or by inflationary fluctuations $\delta h \sim H_{inf}$ which overtop the barrier if $H_{inf} \gtrsim 10^{11}$ GeV \cite{Espinosa:2007qp, EliasMiro:2011aa, Kobakhidze:2013tn}, and this latter problem is especially difficult to avoid if future observations of tensor fluctuations indicate that inflation took place at high scales \cite{Fairbairn:2014zia, Enqvist:2014bua, Kobakhidze:2014xda}.  The stability and the precise bounds on the Hubble scale may also be affected by couplings to gravity \cite{Herranen:2014cua, Kohri:2016wof}, and also on possible couplings of the Higgs field to the inflaton (see also e.g. \cite{Grobov:2015ooa})   Resonant instabilities during preheating and reheating may also destabilize the electroweak vacuum \cite{Gross:2015bea, Ema:2016kpf, Kohri:2016wof, Enqvist:2016mqj}, which further constrains the potential couplings between the Higgs and the inflaton.  Constraints on the Higgs-gravity $\xi \mathcal{H}^{\dagger}\mathcal{H} R$ coupling coming from stability during the postinflationary epoch were discussed in \cite{Herranen:2015ima, Figueroa:2017slm}.  

If the vacuum does decay, given that the present vacuum energy is very slightly positive, the region inside the bubbles will be anti-de-Sitter.  In \cite{Hook:2014uia, Kearney:2015vba, Espinosa:2015qea, East:2016anr} the consequences of the decay to AdS regions for the global structure of the vacuum during and after inflation were investigated: while the global structure of the universe during inflation depends on both the rate of expansion of the inflating regions and the probability of ending up in the terminal AdS vacua, and while the inflating patches may outnumber the AdS regions during and just after inflation, once inflation ends the bubbles may destabilize the electroweak vacuum and consume the entire space if the bubbles do not crunch before they have time to expand.
The exact nature of the decay, and whether it is dominated by quantum, thermal or stochastic processes, depends on the relative scale of $H_{inf}$ as well as the thickness and height of the barrier \cite{Espinosa:2007qp, Hook:2014uia, Espinosa:2015qea}.  Furthermore, it is possible that primordial black holes can serve to nucleate the decay of the electroweak vacuum; this possibility was investigated in \cite{Burda:2015isa, Burda:2016mou}.

The existence of the instability may have observable consequences even if the vacuum remains stable throughout the history of our present observer patch.  The presence of an inflection point in the shape of the metastable Higgs potential may amplify the perturbations of the Higgs field, possibly giving rise to primordial black holes \cite{Espinosa:2017sgp, Espinosa:2018euj} or producing gravity waves \cite{Espinosa:2018eve}.  Alternatively, the stability of the Higgs vacuum during inflation and reheating may indicate the presence of new physics that alters the form of the RG-improved Higgs potential $\lambda(h)h^4/4$ at large values of $h$: in \cite{Bezrukov:2014ipa} temperature-dependent corrections were used to stabilize the electroweak vacuum during inflation and reheating, and in \cite{Lebedev:2012sy} a $(g/2) \phi^2 h^2$ coupling of the Higgs to the inflaton stabilizes the electroweak vacuum during inflation.  In \cite{Hertzberg:2019prp} the superrenormalizable coupling $\phi \mathcal{H}^{\dagger}\mathcal{H}$ is used to assist with stability and also favor the decay of the inflaton into the SM over possible hidden sectors.  Corrections to the Higgs potential may also arise through gravitational effects, giving rise to terms such as $V(\phi)h^2/M_P^2$ or  $h^6/M_P^2$ \cite{Lebedev:2012sy, Hook:2014uia}, and through gravitational couplings to heavy moduli \cite{Ema:2015ehh}.  Other dynamical mechanisms that have been proposed to ensure that the Higgs ends up in the electroweak vacuum include using time-dependent couplings to make the potential barrier stronger in the past \cite{Han:2018yrk}, coupling the Higgs to an additional spectator field \cite{Gong:2017mwt}, and in \cite{Hertzberg:2012zc} a Peccei-Quinn type mechanism is used both to stabilize the Higgs and to provide an axion-like field as a dark matter candidate, in which case there may be a correlation between the Higgs mass and dark matter abundance.  

Note that a negative value for $\lambda(\mu)$ would also affect the form of the potential for the specific Higgs inflation model described in Equation \eqref{HiggsInflation}.  The model can be rescued, however, either if corrections to the potential (such as may arise from the UV completion) at high energies restore $\lambda > 0$, and high-temperature corrections to the potential may keep the electroweak vacuum stable during reheating \cite{Bezrukov:2014ipa}.  If inflation takes place with $\lambda(\mu) > 0$ near but slightly below its critical point $\mu \lesssim \mu_0$, so that $\lambda$ is close to zero, Higgs inflation may be successfully realized for a much lower value of $\xi$,\footnote{Recall from \S 2.1 that $\xi \approx 49000\sqrt{\lambda}$ is necessary to reproduce the observed amplitude of primordial fluctuations.} and the predictions for precision observables may be substantially altered\cite{Hamada:2014iga, Bezrukov:2014bra, Hamada:2014wna}.  This scenario may also give rise to features in the power spectrum, and even lead to the generation of primordial black holes \cite{Ezquiaga:2017fvi, Bezrukov:2017dyv, Rasanen:2018fom}.  

In moduli stabilization models descending from string compactifications it is noted that all known moduli stabilization models uplifting to a de Sitter vacuum are at most metastable rather than absolutely stable (see for instance \cite{Silverstein:2004id, Douglas:2006es}).  Since our own vacuum energy appears to be positive it is therefore maybe not surprising that it should be metastable, but what is perhaps surprising that the metastability should be visible already in the Standard Model, and that the SM should lie so close to the edge of stability.  In \cite{Espinosa:2015qea} it is suggested that the metastability of the Higgs potential may in fact be the dominant effect that limits the lifetime of the present De Sitter phase, in which case quantum gravity effects may somehow be ultimately responsible for fixing the Higgs and top quark masses within a narrow window.  However, it is not a priori clear that the Higgs potential should be related to the cosmological constant problem in this way, since these arise at very different energy scales; furthermore, we reiterate again that given the large number of moduli available in a generic compactification, it is not clear whether we should expect that the Higgs field should be the only field involved in determining the decay rate.

\subsection{Electroweak phase transition}

A much less catastrophic (but still consequential) metastability may arise at low energies, during the electroweak phase transition \cite{Kuzmin:1985mm, Cohen:1991iu, Turok:1990in, Farrar:1993hn}.  The phase transition arises in the Higgs potential in Equation \eqref{HiggsPotentialSM} once thermal corrections are included:
\begin{equation}
V(h, T) = \frac{1}{2}m^2(T) h^2 - \frac{g_3(T)}{3}h^3  + \frac{\lambda(T)}{4}h^4 
\end{equation}
The most important correction term is the cubic term, which can be generated from the thermal terms
\begin{equation}
- \frac{g_3(T)}{3}h^3 \subset -\frac{T}{12\pi} \sum_{i} (m_{i}^2 (h))^{3/2}\,.
\end{equation}
where the index runs over all bosonic degrees of freedom.  For appropriate values of particle masses and couplings, the transition can be first-order, in which case the subsequent departure from equilibrium during vacuum decay and bubble collisions may allow electroweak baryogenesis to occur.  However, the measured values for the Higgs and other particle masses in the SM indicate that the transition is second-order.  Furthermore, the CP violation in the Standard Model is too small for electroweak baryogenesis to successfully reproduce the observed baryon asymmetry.

Several extensions of the SM have been proposed to make the mechanism viable, and these usually require additional particles at or around the TeV scale to make the electroweak transition first-order; see e.g. \cite{Cline:2017jvp} for recent work on the subject.  In this case cosmological observables coming from bubble formation and collisions can serve as a cross-check on the transition: these observables may include generation of gravitational waves \cite{Grojean:2006bp, Addazi:2018nzm, Ellis:2020awk} and of primordial magnetic fields \cite{Baym:1995fk}.  We emphasize, however, that since the electroweak phase transition occurs at energies near those of particle physics, further evidence of new particles at the TeV scale may be required as well in order to confirm that the phase transition is a consequence of the Higgs field.  

\section{The Higgs field and the landscape}

Can the Higgs field help us understand whether our low-energy vacuum is part of a larger landscape, and what corner of it we inhabit?  Conversely, we might ask whether top-down considerations from string theory or field theory can lead to constraints or predictions for models of low-energy Higgs physics.  In addition to these general considerations, specific models of moduli dynamics within the landscape can also help suggest dynamical mechanisms with interesting consequences for Higgs physics.

Moduli fields are a general feature of supersymmetric field theories and string compactifications, where they can arise from the shapes and sizes of the compactification manifold and its subcycles, as well as higher-dimensional forms, fields and sources living there.  The variety of different types of moduli that can arise, as well as the many different techniques for their stabilization (see for instance \cite{Silverstein:2004id, Douglas:2006es}), is helpful for constructing realizations of specific low-energy particle physics models in a UV-complete framework, but on the other hand the array of options available limits the top-down predictivity of the larger framework.  Nevertheless, a few general statements can be made: one is that the existence of many moduli, which must be stabilized in order to avoid appearing in the low-energy vacuum, seems difficult to avoid.  A second is that although supersymmetry leads to powerful calculational techniques in the context of string and field theory, and worldsheet supersymmetry is an important part of many formulations of string theory, spacetime supersymmetry at the electroweak scale is not guaranteed (see e.g. \cite{Douglas:2006es, Silverstein:2016ggb}).  A third is that specific examples which realize the Standard Model with $m_h = 125$ GeV within the context of a fully worked model with all moduli fixed can be found (see e.g.\ \cite{Kane:2011kj}), though it is not clear that these are enough to determine which model of moduli stabilization may be realized in our universe.

Some of the strongest constraints on moduli and their stabilization come from cosmology: the cosmological moduli problem \cite{Coughlan:1983ci, deCarlos:1993wie, Banks:1993en} consists of the observation that the presence of heavy moduli fields may overclose the universe.  Fields $\phi$ displaced from the minimum of their potential will remain frozen by Hubble friction until $m_{\phi} \sim H$, at which point they will begin to oscillate with $\rho \propto a(t)^{-3}$.  If these decay to lighter particles or radiation through gravitational strength couplings their decay rate can be estimated as
\begin{equation}
\Gamma \sim \frac{m_{\phi}^{3}}{M_P^2}\,.
\end{equation}
In order to avoid spoiling the predictions of Big Bang Nucleosynthesis we require $H \sim \Gamma \gtrsim 1 MeV$ at the time of moduli decay, which implies that moduli must be stabilized with $m_{\phi} \gtrsim 30$ TeV.

Moduli fields that oscillate for awhile between a first and second period of reheating may also give rise to a non-thermal history where a period of matter domination precedes the hot Big Bang phase (see e.g. \cite{Kane:2015jia, Allahverdi:2020bys} for a review).  This was used in \cite{Hardy:2018bph, Bernal:2018ins} as a way to relax the constraints on Higgs portal dark matter.  Such models could also be complemented by searches for rare Higgs decays at the high-intensity LHC, or indirectly through astrophysical dark matter searches.

Even heavy moduli fields can have important consequences for the low-energy theory, however: for instance, the potential of light fields may be affected by the backreaction and rearrangement of heavy fields \cite{Dong:2010in}\footnote{As mentioned above, the model of Higgs inflation described in \cite{Bezrukov:2007ep} can be considered an example of this effect.}.  Another effect from beyond the regime of validity of perturbative field theory is particle production and moduli trapping at enhanced symmetry points \cite{Kofman:2004yc, Watson:2004aq}.  This may arise in string theory models when two branes collide and strings connecting them are formed and prevent further motion, in what may be considered a stringy version of the Higgs mechanism.  This phenomenon of moduli trapping may provide a dynamical selection mechanism for trapping the evolution of scalar fields at points of enhanced symmetry in the low-energy field theory landscape, perhaps helping to explain the high degree of symmetry found in the Standard Model.

Many models of moduli stabilization feature light moduli coming from flat directions with a shift symmetry in the potential -- these may arise from higher-form fluxes wrapping the extra dimensions, which give rise to axion-like fields in the low energy theory.  Instantons break the shift symmetry nonperturbatively to a discrete shift symmetry (i.e., corresponding to a periodic instead of a flat potential) and give the axions a mass; since the mass depends exponentially on the instanton action, the possible range of values of axion masses may span many orders of magnitude.  (For explorations of cosmology with many ultralight axions distributed across different mass scales, see e.g. \cite{Arvanitaki:2009fg, Hui:2016ltb}.)  The axion may also have its potential lifted by nonperiodic effects; this axion monodromy has been used to develop large field models of inflation in string compactifications \cite{Silverstein:2008sg, McAllister:2008hb}.  Flat axion-like directions have also been applied to search for dynamical mechanisms to stabilize the Higgs mass: the relaxion mechanism \cite{Graham:2015cka} uses both the shift symmetry and its nonperturbative breaking to stabilize the Higgs mass as follows: in the simplest version of this class of models the SM Lagrangian is assumed to contain the extra terms
\begin{equation}
\mathcal{L}_{SM} = (-M^2 + g \phi)|h|^2 + V(g\phi) + \frac{1}{32\pi^2}\frac{\phi}{f}\tilde{G}_{\mu \nu}G^{\mu \nu}
\end{equation} 
where $\phi$ is the inflaton, $f$ is the axion decay constant, $g$ sets the scale of shift-symmetry breaking, $M$ is the UV cutoff, and $G_{\mu \nu}$ is the QCD field strength.  Below the QCD scale $\Lambda_{QCD}$, the effective field theory becomes
\begin{equation}
(-M^2 + g\phi) + (gM^2 \phi + g^2 \phi^2 + ...) + \Lambda^4 \cos(\phi/f)
\end{equation} 
where the ellipses indicate radiative corrections of higher order in $g\phi/M^2$, and $\Lambda$ is determined by $\Lambda_{QCD}$ and by the quark masses.  As the Higgs mass squared becomes negative, the symmetry breaks and the Higgs field gets a vev, the quark masses increase and increase the barrier heights in the oscillating potential, preventing further rolling of $\phi$.  This model therefore provides a dynamical mechanism for making the smallness of the electroweak scale technically natural, since it depends on the smallness of the symmetry-breaking parameters $g$ and $\Lambda_{QCD}$.  (The cutoff scale for this class of models may not be as high as the Planck scale; however, there may be further physics to preserve naturalness at the cutoff.)  Cosmological constraints on this mechanism may be model-dependent but may be affected by cosmological constraints on axion dark matter or long-range forces.  The simplest versions of the model also call for new physics (or a hidden sector) around the electroweak scale, which may be accessible in either dark matter or collider searches.  
A similar class of mechanisms involves coupling the Higgs field instead to higher-dimensional form flux backgrounds \cite{Giudice:2019iwl, Kaloper:2019xfj} via terms such as $\frac{g}{24}\epsilon^{\mu \nu \rho \sigma}F_{\mu \nu \rho \sigma} |h|^2$, which decay via nonperturbative brane nucleation\footnote{Note that the four-form field is related to an axion by Poincar\'e duality in four dimensions.}.  See also \cite{Geller:2018xvz} for a model where inflation is used to select a weak scale corresponding to the maximum potential of an axion-like modulus.

Cosmological relaxation has also been proposed for fixing the size of the cosmological constant \cite{Graham:2019bfu} at much lower energy scales, and the authors of \cite{Giudice:2019iwl, Kaloper:2019xfj} also discuss whether both the Higgs mass and the electroweak and cosmological constant can be selected by the same dynamical mechanism.  It is difficult, however, to generate the hierarchy between the electroweak scale and the cosmological constant using the same physics, and anthropic arguments may need to be taken into consideration.

\section{Conclusions and future directions}

In this brief review we have only scratched the surface of a rich and extensive subject, and hopefully the literature discussed here will help guide the reader to further reading on the Higgs field and cosmology.  The Higgs field can play a variety of roles in the early universe, either as a single scalar field or as part of a dynamic and complex scalar sector, and cosmological constraints and observables have the potential to teach us a great deal about the dynamics of the Higgs field.  
To summarize a few of the specific points discussed in this review:

\begin{itemize}

\item While it is possible to build models where the Standard Model Higgs field serves as the inflaton, generating the observed primordial power spectrum requires the quartic Higgs potential to be flattened somehow at large field values.  Coupling the Higgs to gravity can accomplish this, but the existence of nonrenormalizable corrections to the field theory means that this class of models may not be as simple to realize, nor its predictions as specific and universal, as originally envisioned.

\item Even when the Higgs field makes a subdominant contribution to the energy during inflation, its dynamics may be imprinted on the primordial power spectrum via one of several different mechanisms (e.g. curvaton, modulated reheating, or ``Higgscitement'').  Precision observables such as the non-Gaussianities, the spectral tilt and the tensor-to-scalar ratio can provide strong but model-dependent constraints on such scenarios.  Furthermore, the fact that the Higgs couplings to the Standard Model are well understood makes the Higgs useful for building models of reheating and the exit from inflation.

\item The measured values of the Standard Model parameters (and particularly the Higgs mass and the top quark Yukawa coupling) are consistent with a metastable Standard Model vacuum that lies close to the edge of stability.  Ensuring that the Universe ends up in the electroweak vacuum after inflation and reheating therefore provides a strong constraint on model building.  The fact that the SM is on the edge of stability may have a dynamical or anthropic explanation, or a combination thereof. 

\item The electroweak phase transition will not be first-order, nor will CP violation be sufficiently large for electroweak baryogenesis to occur, without extensions to the Standard Model.  If extensions to the SM do result in electroweak baryogenesis, however, cosmological signatures from the phase transition can be cross-correlated with observations of new particles and couplings slightly at or above the TeV scale.

\item  Although it is difficult to predict where the SM vacuum may lie in the landscape, the genericity of models with additional moduli raises the possibility that the Higgs couples to additional scalar moduli with interesting cosmological consequences.  Understanding Higgs and moduli dynamics in the early universe may therefore help us understand our place in the landscape, and whether it is dynamically or anthropically selected.  
However, the cosmological constant problem still looms, and it is not clear that the same dynamics can resolve both the cosmological constant problem and the hierarchy problem simultaneously without resorting to anthropic arguments.

\item Nonperturbative dynamics of the Higgs field and other scalar fields may imprint experimental cosmological signatures such as features in the primordial power spectrum, stochastic gravity wave backgrounds, or primordial magnetic fields.  

\end{itemize}

In the coming years, measurements of the Higgs from the energy, intensity and cosmic frontiers will combine to help us to identify and explore our corner of the landscape of possible models.
We hope that these studies will help us to understand the origin and fate of our Universe, and help us understand the role that the concept of naturalness may play as a guiding principle.

\section*{Acknowledgments}

We thank Daniel Figueroa, Rostislav Konoplich, Eva Silverstein and Alex Westphal for very helpful discussions.  This research was supported in part by the Office of the Provost and the Office of the Dean of the School of Science at Manhattan College.


\end{document}